\documentclass [12 pt]{article}
\def\be{\begin{equation}}
\def\eea{\end{eqnarray}}
\def\bea{\begin{eqnarray}}
\def\ee{\end{equation}}

\usepackage[psamsfonts]{amssymb}
\usepackage{amsmath}
\author{F. Kheirandish$^{1}$ \footnote{fardin$_{-}$kh@phys.ui.ac.ir} and M.
Amooshahi$^{1}$ \footnote{amooshahi@sci.ui.ac.ir}
\\ $^{1}$ {\small Department of Physics, University of Isfahan,}
\\ {\small Hezar Jarib Ave., Isfahan, Iran.}}
\title{Electromagnetic field quantization in a linear dielectric medium}
\begin{document}
\maketitle
\begin{abstract}
\noindent By modeling a dielectric medium with two independent
reservoirs, i.e., electric and magnetic reservoirs, the
electromagnetic field is quantized in a linear dielectric medium
consistently. A Hamiltonian is proposed from which using the
Heisenberg equations, not only the Maxwell equations but also the
structural equations can be obtained. Using the Laplace
transformation, the wave equation for the electromagnetic vector
potential is solved in the case of a homogeneous dielectric
medium. Some examples are considered showing the applicability of
the model to both absorptive and nonabsorptive dielectrics.
\end{abstract}
\section{Introduction}
In a homogeneous and nondispersive medium, the photon is
associated with only the transverse part of the electromagnetic
field, in contrast in an inhomogeneous nondispersive medium, the
transverse and the longitudinal degrees of freedom are coupled.
In this case the quantization of the electromagnetic field can be
accomplished by employing a generalized gauge [1,2], that is,
$\vec{\nabla}\cdot(\varepsilon(\vec{r})\vec{A})=0$, where
$\varepsilon(\vec{r})$ is the space dependent dielectric
function. Using the gauge $\sum_{j=1}^3\frac{\partial}{\partial
x_j}(\varepsilon_{ij}(\vec{r})\vec{A}_j)=0$ [3], the
generalization of this quantization to the case of an anisotropic
nondispersive medium is straightforward. \\
The quantization in a dispersive and absorptive dielectric,
represents one of the most interesting problems in quantum
optics, because it gives a rigorous test of our understanding of
the interaction of light with matter. The dissipative nature of a
medium is an immediate consequence of its disspersive character
and vice versa according to the Kramers-Kronig relations. This
means that the validities of the electromagnetic quantization in
nondissipative but dispersive media is restricted to some range of
frequencies for which the imaginary part of the dielectric
function is negligible, otherwise there will be
inconsistencies.\\
In the scheme of Lenac [4], for dispersive and nonabsorptive
dielectric media, by starting with fundamental equations of motion
, the medium is described by a dielectric function $
\varepsilon(\vec{r},\omega)$, without any restriction on its
spatial behavior. In this scheme it is assumed there is no losses
in the system, so the dielectric function is real for the whole
space. The procedure is based on an expansion of the total field
in terms of the coupled eigenmodes, orthogonality relations are
derived and equal-time commutation relations are discussed.\\
Huttner and Barnett [5] have presented a canonical quantization
for electromagnetic field inside the dispersive and absorptive
dieletrics based on a microscopic model in which the medium is
represented by a collection of interacting matter fields and the
absorptive character of the medium is described by interaction of
the matter fields with a reservoir containing a continuum of the
Klein-Gordon fields. In their model, eigen-operators for coupled
systems are calculated and electromagnetic field has been
expressed in terms of them, the dielectric function is derived and
it is shown to satisfy the Kramers-Kronig relations.\\
Gruner and Welsch [6], presented a quantization method of the
radiation field inside a dispersive and absorptive linear
dielectric starting from the phenomenological Maxwell equations,
where the properties of the dielectric are described by a
permitivity consistent with the Kramers-Kronig relations, an
expansion of the field operators is performed that is based on
the Green function of the classical
Maxwell equations which preserves the equal-time canonical field commutation relations.\\
Suttorp and Wubs [8] in the framework of a damped polarization
model have quantized the electromagnetic field in an absorptive
medium with spatial dependendence of its parameters. They have
solved the equations of motion of the dielectric polarization and
the eletromagnetic field by means of the Laplace transformation
for both positive and negative times. The operators that
diagonalize the Hamiltonian are found as linear combinations of
canonical variables with coefficients depending on the electric
susceptibility and the dielectric Green function. Also the time
dependence of the electromagnetic field and the dielectric
polarization are determined.\\
The macroscopic description of a quantum damped harmonic
oscillator is presented in terms of Langevin equation [9,10]. The
coupling with the heath bath in the microscopic theory
corresponds to two terms in the macroscopic equation of motion of
a damped harmonic oscillator, a random force and a memory function
as
\begin{equation}\label{I2}
\vec{\ddot{x}}+\int_0^\infty
dt'\mu(t-t')\vec{\dot{x}}(t')-\omega_0^2\int_0^\infty
dt'\nu(t-t')\vec{x}(t')=F_N(t),
\end{equation}
where $ F_N(t) $ is the random force and $ \omega_0 $ is the
frequency of the oscillator. Matloob [11] quantized the
macroscopic electromagnetic field in a linear isotropic permeable
dielectric medium by quantizing the Langevin equation and
associating a damped quantum harmonic oscillator with each mode
of the radiation field. There are some another approaches for
quantizing the
electromagnetic field, for example see [12-22].\\
The structure of this article is as follows: \\
In section 2, we quantize the electromagnetic field in a linear
 dielectric medium by modeling
 the dielectric medium with two independant reservoirs, i.e.,
magnetic and electric reservoirs respectively, where the electric
reservoir interacts with the displacement vector field through a
minimal coupling term and the magnetic reservoir interacts with
the magnetic field. It is shown that both the Maxwell and
structural equations can be obtained from Heisenberg equations.
In section 3, we solve the equations of motion for a homogeneous
dielectric medium using the Laplace transformation. Finally we
give some examples indicating the applicability of the model to
the case of absorptive or nonabsorptive dielectrics.
\section{ Quantum dynamics}
Quantum electrodynamics in a linear dielectric medium can be
investigated by introducing two reservoirs that interact with the
electromagnetic field  through a new kind of minimal coupling.
The Heisenberg equations for eletromagnetic field and the
reservoirs give not only the Maxwell equations but also the
structural equations which relate the electric and magnetic fields
to the electric and magnetic polarization densities respectively.
The vector potential of the electromagnetic field can be written
in terms of plane waves
\begin{equation} \label{d1}
\vec{A}(\vec{r},t)=\int d^3 \vec{k} \sum_{\lambda=1}^2
\sqrt{\frac{\hbar}
{2(2\pi)^3\varepsilon_0\omega_{\vec{k}}}}[a_{\vec{k}\lambda}(t)e^{i\vec{k}.\vec{r}}+a_{\vec{k}
\lambda}^\dag(t)e^{-i\vec{k}.\vec{r}}]\vec{e}_{\vec{k}\lambda}
\end{equation}
where $ \omega_{\vec{k}}=c|\vec{k}| $ , $\varepsilon_0 $ is the
permitivity of the vacuum and
 $\vec{e}_{\vec{k}\lambda}, \hspace{00.50 cm} (\lambda=1,2) $ are
polarization vectors which satisfy
\begin{eqnarray}\label{d1.5}
\vec{e}_{\vec{k}\lambda}.\vec{e}_{\vec{k}\lambda'}&=&\delta_{\lambda\lambda'},\nonumber\\
 \vec{e}_{\vec{k}\lambda}.\vec{k}&=&0,
\end{eqnarray}
These recent relations guarantee that the vector potential in
(\ref{d1}) satisfies the Coulomb gauge $
\vec{\nabla}\cdot\vec{A}(\vec{r},t)=0 $.\\
 Operators $a_{\vec{k}\lambda}(t)$ and $a_{\vec{k}\lambda}^\dag(t)
$ are annihilation and creation operators of the electromagnetic
field and satisfy the following equal time commutation rules
\begin{equation}\label{d2}
[a_{\vec{k}\lambda}(t),a_{\vec{k'}\lambda'}^\dag(t)]=
\delta(\vec{k}-\vec{k'})\delta_{\lambda\lambda'}.
\end{equation}
The conjugate canonical momentum density of the electromagnetic
field $\vec{\pi}_F(\vec{r},t)$ and also the displacement vector
operator $\vec{D}(\vec{r},t)$, are by definition
\begin{equation}\label{d3}
\vec{\pi}_F(\vec{r},t)=- \vec{D}(\vec{r},t)=i\varepsilon_0\int
d^3\vec{k} \sum_{\lambda=1}^2
\sqrt{\frac{\hbar\omega_{\vec{k}}}{2(2\pi)^3\varepsilon_0}}
[a_{\vec{k}\lambda}^\dag(t)e^{-i\vec{k}.\vec{r}}-a_{\vec{k}\lambda}(t)e^{i\vec{k}.\vec{r}}]\vec{e}_{\vec{k}\lambda}.
\end{equation}
From this definition it is obvious that $ \nabla\cdot \vec{D}=0 $
which is the Gauss law in the absence of external charges. The
commutation relations (\ref{d2}) lead to the commutation relations
between the components of the vector potential $\vec{A}$ and the
displacement vector operator $\vec{D}$
\begin{equation}\label{d3.1}
[A_l(\vec{r},t),-D_j(\vec{r'},t)]=[A_l(\vec{r},t),\pi_j(\vec{r'},t)]=
\imath\hbar\delta_{lj}^\bot(\vec{r}-\vec{r'}),
\end{equation}
where $\delta_{lj}^\bot(\vec{r}-\vec{r'})=\frac{1}{(2\pi)^3}\int
d^3\vec{k}e^{i\vec{k}\cdot(\vec{r}-\vec{r'})}(\delta_{lj}-\frac{k_l
k_j }{|\vec{k}|^2})$, is the transverse delta function
with the following properties. \\
\hspace{1cm}
 1. Let the transverse $\vec{F}^\bot(\vec{r},t)$ and the
 longitudinal $\vec{F}^\|(\vec{r},t)$ coponents of an arbitrary vector field $
\vec{F}(\vec{r},t) $ be defined as
\begin{eqnarray}\label{d3.3}
&& \vec{F}^\bot(\vec{r},t)=\vec{F}(\vec{r},t)-\int d^3r'
\nabla'\cdot\vec{F}(\vec{r'},t)\vec{\nabla} G(\vec{r},\vec{r'}), \\
&&\vec{F}^\|(\vec{r},t)=\int d^3 r'
\nabla'\cdot\vec{F}(\vec{r'},t)\vec{\nabla} G(\vec{r},\vec{r'}),
\end{eqnarray}
where $G(\vec{r},\vec{r'})$, is the Green function
\begin{equation}\label{d3.4}
G(\vec{r},\vec{r'})=-\frac{1}{4\pi|\vec{r}-\vec{r'}|},
\end{equation}
then as the first property, one can easily show that
\begin{equation}\label{d3.5}
F_i^\bot(\vec{r},t)=\sum_{j=1}^3\int
d^3r'\delta_{ij}^\bot(\vec{r}-\vec{r'})F_j(\vec{r'}).
\end{equation}
2. The second property is
\begin{equation}\label{d3.6}
\sum_{l=1}^3\frac{\partial}{\partial
x_l}\delta_{lj}^\bot(\vec{r}-\vec{r'})=0,\hspace{1.00 cm}j=1,2,3.
\end{equation}
The Hamiltonian of the electromagnetic field inside the
dielectric can be defined as
\begin{eqnarray}\label{d4.5}
H_F(t)&=&\int d^3r [\frac{ \vec{D}^2(\vec{r},t)}{2\varepsilon_0}+
\frac{(\nabla\times\vec{A})^2(\vec{r},t)}{2\mu_0}],\nonumber\\
&=&\sum_{\lambda=1}^2\int d^3\vec{k}\hbar
\omega_{\vec{k}}[a_{\vec{k}\lambda}^\dag(t)
a_{\vec{k}\lambda}(t)+\frac{1}{2}],
\end{eqnarray}
where $\mu_0$ is the magnetic permitivity of the vacuum.\\ Now we
 propose the total Hamiltonian, i.e., electromagnetic field plus
reservoirs like this
\begin{eqnarray}\label{d4.55}
\tilde{H}(t)&=&\int d^3r \{\frac{[
\vec{D}(\vec{r},t)-\vec{P}(\vec{r},t)]^2}{2\varepsilon_0}+
\frac{(\nabla\times\vec{A})^2(\vec{r},t)}{2\mu_0}
-\nabla\times\vec{A}(\vec(\vec{r},t).\vec{M}(\vec{r},t)\}\nonumber\\
&+&+H_e+H_m,
\end{eqnarray}
where $H_e$ and $H_m$ are the Hamiltonians of the electric and
magnetic reservoirs respectively
\begin{equation}\label{d4.6}
H_e(t)=\sum_{\nu=1}^3\int d^3\vec{q}\int d^3\vec{k}
\hbar\omega_{\vec{k}}
[d_{\nu}^\dag(\vec{k},\vec{q},t)d_{\nu}(\vec{k},\vec{q},t)
+\frac{1}{2}],
\end{equation}
\begin{equation}\label{d4.61}
H_m(t)=\sum_{\nu=1}^3\int d^3\vec{q}\int d^3\vec{k}
\hbar\omega_{\vec{k}}
[b_{\nu}^\dag(\vec{k},\vec{q},t)b_{\nu}(\vec{k},\vec{q},t)
+\frac{1}{2}].
\end{equation}
The annihilation and creation operators
$d_{\nu}(\vec{k},\vec{q},t)$, $d_{\nu}^\dag(\vec{k},\vec{q},t)$,
$b_{\nu}(\vec{k},\vec{q},t)$ and
$b_{\nu}^\dag(\vec{k},\vec{q},t)$ of the electric and magnetic
reservoirs respectively, satisfy the following equal time
commutation relations
\begin{eqnarray}\label{d4.7}
&&[d_{\nu}(\vec{k},\vec{q},t) ,
d_{\nu'}^\dag(\vec{k'},\vec{q'},t)]=
\delta_{\nu\nu'}\delta(\vec{k}-\vec{k}')\delta(\vec{q}-\vec{q}'),\nonumber\\
&&[b_{\nu}(\vec{k},\vec{q},t),b_{\nu'}^\dag(\vec{k'},\vec{q'},t)]=
\delta_{\nu\nu'}\delta(\vec{k}-\vec{k}')\delta(\vec{q}-\vec{q}').
\end{eqnarray}
Accordingly the electric and magnetic polarization density
operators $\vec{P}(\vec{r},t)$ and $ \vec{M}(\vec{r},t)$, in
(\ref{d4.55}), can be expanded as
\begin{eqnarray}\label{d4.72}
&&\vec{P}(\vec{r},t)=\sum_{\nu=1}^3 \int
\frac{d^3\vec{q}}{\sqrt{(2\pi)^3}} \int
d^3\vec{k}[f(\omega_{\vec{k}},\vec{r})d_{\nu}(\vec{k},\vec{q},t)e^{i\vec{q}.\vec{r}}+
f^*(\omega_{\vec{k}},\vec{r})d_{\nu}^\dag(\vec{k},\vec{q},t)e^{-i\vec{q}.\vec{r}}]
\vec{v}_{\nu}(\vec{q})\nonumber\\
&&
\end{eqnarray}
\begin{eqnarray}\label{d4.73}
&& \vec{M}(\vec{r},t)=i\sum_{\nu=1}^3 \int
\frac{d^3\vec{q}}{\sqrt{(2\pi)^3}}\int
d^3\vec{k}[g(\omega_{\vec{k}},\vec{r})b_{\nu}(\vec{k},\vec{q},t)e^{i\vec{q}.\vec{r}}-
g^*(\omega_{\vec{k}},\vec{r})b_{\nu}^\dag(\vec{k},\vec{q},t)e^{-i\vec{q}.\vec{r}}]\vec{s}_{\nu}(\vec{q}),\nonumber\\
&&
\end{eqnarray}
where
\begin{eqnarray}\label{d5.1}
\vec{v}_{\nu}(\vec{q})&=&\vec{e}_{\vec{k}\nu}(\vec{r}),\hspace{1cm}\nu=1,2\nonumber\\
\vec{v}_{3}(\vec{q})&=&\hat{q}=
\frac{\vec{q}}{|\vec{q}|},\nonumber\\
\vec{s}_{\nu}(\vec{q})&=&\hat{q}\times\vec{e}_{\nu\vec{q}}
,\hspace{1cm}\nu=1,2,\nonumber\\
\vec{s}_{3}(\vec{q})&=&\hat{q}.
\end{eqnarray}
 In definition of polarization densities (\ref{d4.72}), the functions
 $ f(\omega_{\vec{k}},\vec{r})$ and $g(\omega_{\vec{k}},\vec{r}) $, are the coupling
 functions between the electromagnetic field and the electric and magnetic reservoirs
 respectively and are dependent on position $ \vec{r} $ for
 inhomogeneous dielectrics.
Using (\ref{d3.1}) and (\ref{d3.5}), we can obtain the Heisenberg
equations for $\vec{A}$ and $\vec{D}$
\begin{equation}\label{d6}
\frac{\partial\vec{A}(\vec{r},t)}{\partial
t}=\frac{\imath}{\hbar}[\tilde{H},\vec{A}(\vec{r},t)]=
-\frac{\vec{D}(\vec{r},t)-\vec{P}^\bot(\vec{r},t)}{\varepsilon_0},
\end{equation}
\begin{equation}\label{d6.1}
\frac{\partial\vec{D}(\vec{r},t)}{\partial
t}=\frac{\imath}{\hbar}[\tilde{H},\vec{D}(\vec{r},t)]=
\frac{\nabla\times\nabla\times\vec{A}(\vec{r},t)}{\mu_0}-\nabla\times\vec{M}^\bot(\vec{r},t),
\end{equation}
where $\vec{P}^\bot$ and $\vec{M}^\bot$ are transverse components
of $\vec{P}$ and $\vec{M}$. Now define transverse electrical
field $ \vec{E}^\bot $, magnetic field $\vec{B}$ and
 magnetic induction field  $ \vec{H} $ as
 \begin{equation}\label{d7}
 \vec{E}^\bot=-\frac{\partial\vec{A}}{\partial t},\hspace{1.00
 cm}\vec{B}=\nabla\times\vec{A},\hspace{1.00
 cm}\vec{H}=\frac{\vec{B}}{\mu_0}-\vec{M},
 \end{equation}
 respectively, then (\ref{d6}) and (\ref{d6.1}) can be rewritten as
\begin{equation} \label{d8}
\vec{D}=\varepsilon_0 \vec{E}^\bot+\vec{P}^\bot,
\end{equation}
\begin{equation}\label{d8.1}
\frac{\partial \vec{D}}{\partial t}=\nabla\times\vec{H}^\bot,
\end{equation}
in the absence of external charge density, we have
$\vec{D}^\|=\varepsilon_0\vec{E}^\|+\vec{P}^\|=0$ and we can
define the longitudinal component of the electrical field as
$\vec{E}^\|=-\frac{\vec{P}^\|}{\varepsilon_0}$.\\
 Combination of
(\ref{d6}) and (\ref{d6.1}) leads to
\begin{equation}\label{d9}
-\nabla^2\vec{A}+\frac{1}{c^2}\frac{\partial^2\vec{A}}{\partial
t^2}=\mu_0\frac{\partial\vec{P}^\bot}{\partial
t}+\mu_0\nabla\times\vec{M}^\bot.
\end{equation}
Using commutation relations (\ref{d4.7}) we find the Heisenberg
equations for operators $ d_{\vec{n}\nu}(\vec{k},t) $ and $
b_{\vec{n}\nu}(\vec{k},t)$,
\begin{eqnarray}\label{d10}
&&\dot{d}_{\nu}(\vec{k},\vec{q},t)=
\frac{\imath}{\hbar}[\tilde{H},d_{\nu}(\vec{k},\vec{q},t)]=\nonumber\\
&&-\imath\omega_{\vec{k}}d_{\nu}(\vec{k},\vec{q},t)+\frac{\imath}{\hbar\sqrt{(2\pi)^3}}
\int d^3r'f^*(\omega_{\vec{k}},\vec{r'})
\vec{E}(\vec{r'},t)e^{-i\vec{q}\cdot\vec{r'}}\cdot\vec{v}_{\nu}(\vec{q}),\nonumber\\
&&
\end{eqnarray}
\begin{eqnarray}\label{d11}
&&\dot{b}_{\nu}(\vec{k},\vec{q},t)=\frac{\imath}{\hbar}[\tilde{H},b_{\nu}(\vec{k},\vec{q},t)]=
\nonumber\\
&-&\imath\omega_{\vec{k}}b_{\nu}(\vec{k},\vec{q},t)
+\frac{1}{\hbar\sqrt{(2\pi)^3}} \int
d^3r'g^*(\omega_{\vec{k}},\vec{r'})e^{-i\vec{q}\cdot\vec{r'}}\vec{B}(\vec{r'},t)\cdot
\vec{s}_{\nu}(\vec{q}),
\end{eqnarray}
 with the following formal solutions
\begin{eqnarray}\label{d11.1}
&&{d}_{\nu}(\vec{k},\vec{q},t)=
d_{\nu}(\vec{k},\vec{q},0)e^{-\imath\omega_{\vec{k}}t}+\nonumber\\
&&\frac{\imath}{\hbar\sqrt{(2\pi)^3}}\int_0^t
dt'e^{-\imath\omega_{\vec{k}}(t-t')} \int
d^3r'f^*(\omega_{\vec{k}},\vec{r'})e^{-i\vec{q}\cdot\vec{r'}}
\vec{E}(\vec{r'},t')\cdot\vec{v}_{\nu}(\vec{q}),\nonumber\\
&&
\end{eqnarray}
\begin{eqnarray}\label{d11.2}
&&{b}_{\nu}(\vec{k},\vec{q},t)=
b_{\nu}(\vec{k},\vec{q},0)e^{-\imath\omega_{\vec{k}}t}+\nonumber\\
&&\frac{1}{\hbar\sqrt{(2\pi)^3}}\int_0^t
dt'e^{-\imath\omega_{\vec{k}}(t-t')} \int
d^3r'g^*(\omega_{\vec{k}},\vec{r'})e^{-i\vec{q}\cdot\vec{r'}}\vec{B}
(\vec{r'},t')\cdot\vec{s}_{\nu}(\vec{q}).
\end{eqnarray}
 After substituting of (\ref{d11.1}) in (\ref{d4.72}),
\begin{equation}\label{d12}
\vec{P}(\vec{r},t)=\vec{P}_N(\vec{r},t)+\varepsilon_0\int_0^{|t|}
d t' \chi_e(\vec{r},|t|-t')\vec{E}(\vec{r},\pm t'),
\end{equation}
where the upper (lower) sing corresponds to $t>0$ ($ t<0 $)
respectively. The total electric field is
$\vec{E}=\vec{E}^\bot+\vec{E}^\|=-\frac{\partial\vec{A}}{\partial
t}-\frac{\vec{P}^\|}{\varepsilon_0}$ and the memory function
\begin{eqnarray}\label{d13}
\chi_e(\vec{r},t)&=&\frac{8\pi}{\hbar c^3
\varepsilon_0}\int_0^\infty
d\omega_{\vec{k}}\omega_{\vec{k}}^2|f(\vec{r},\omega_{\vec{k}})|^2\sin\omega_{\vec{k}}t,
\hspace{1cm}t>0,\nonumber\\
\chi_e(\vec{r},t)&=&0,\hspace{1.5cm} t\leq 0,
\end{eqnarray}
is called the electric susceptibility of the dielectric. If we are
given a definite $\chi_e(t)$ which is zero for $t\leq 0 $, then
we can obtain the corresponding coupling function
$f(\vec{r},\omega_{\vec{k}})$,
\begin{eqnarray}\label{d14}
|f(\vec{r},\omega_{\vec{k}})|^2&=&\frac{\hbar c^3 \varepsilon_0
}{4\pi^2\omega_{\vec{k}}^2}\int_0^\infty dt\chi_e(\vec{r},t)
\sin\omega_{\vec{k}}t,\hspace{1cm}\omega_{\vec{k}}>0,\nonumber\\
|f(\vec{r},\omega_{\vec{k}})|^2&=&0,\hspace{1.00cm}\omega_{\vec{k}}=0.
\end{eqnarray}
The operator $\vec{P}_N(\vec{r},t) $ in (\ref{d12}), is the noise
electric polarization density
\begin{eqnarray}\label{d16}
&&\vec{P}_N(\vec{r},t)=\nonumber\\
&&\sum_{\nu=1}^3\int\frac{d^3\vec{q}}{\sqrt{(2\pi)^3}}\int
d^3\vec{k}[f(\omega_{\vec{k}},\vec{r})d_{\nu}(\vec{k},\vec{q},0)
e^{-\imath\omega_{\vec{k}}t+i\vec{q}.\vec{r}}+f^*(\omega_{\vec{k}},\vec{r})
d_{\nu}^\dag(\vec{k},\vec{q},0)e^{\imath\omega_{\vec{k}}t-i\vec{q}.\vec{r}}]\vec{v}_{
\nu}(\vec{q}).\nonumber\\
\end{eqnarray}
Similarly after substituting (\ref{d11.2}) in (\ref{d4.73}), we
obtain the following expression for $\vec{M}(\vec{r},t)$,
\begin{equation}\label{d17}
\vec{M}(\vec{r},t)=\vec{M}_N(\vec{r},t)+\frac{1}{\mu_0}\int_0^{|t|}
dt' \chi_m(\vec{r},|t|-t')\vec{B}(\vec{r},\pm t'),
\end{equation}
where $\chi_m $ is the magnetic susceptibility of the dielectric
\begin{eqnarray}\label{d19}
\chi_m(\vec{r},t)&=&\frac{8\pi\mu_0}{\hbar c^3}\int_0^\infty
d\omega_{\vec{k}}\omega_{\vec{k}}^2|g(\vec{r},\omega_{\vec{k}})|^2
\sin\omega_{\vec{k}}t,\hspace{1cm}t>0,\nonumber\\
\chi_m(\vec{r},t)&=&0,\hspace{1.5cm}t\leq 0.
\end{eqnarray}
If we are given a definite $\chi_m(t)$, then we can obtain the
corresponding coupling function $g(\vec{r},\omega_{\vec{k}})$ in
terms of $\chi_m(\vec{r},t)$
\begin{eqnarray}\label{d20}
|g(\vec{r},\omega_{\vec{k}})|^2&=&\frac{\hbar c^3
}{4\pi^2\mu_0\omega_{\vec{k}}^2}\int_0^\infty dt\chi_m(\vec{r},t)
\sin\omega_{\vec{k}}t,\hspace{1cm}\omega_{\vec{k}}>0,\nonumber\\
|g(\vec{r},\omega_{\vec{k}})|^2&=&0,\hspace{1cm}\omega_{\vec{k}}=0.
\end{eqnarray}
The operator $M_N(\vec{r},t)$ is the noise magnetic dipole
density
\begin{eqnarray}\label{d22}
&&\vec{M}_N(\vec{r},t)=\nonumber\\
&&i\sum_{\nu=1}^3\int\int\frac{d^3\vec{q}}{\sqrt{(2\pi)^3}}
d^3\vec{k}[g(\omega_{\vec{k}},\vec{r})b_{\nu}(\vec{k},\vec{q},0)
e^{-\imath\omega_{\vec{k}}t+i\vec{q}\cdot\vec{r}}-g^*(\omega_{\vec{k}},\vec{r})
b_{\nu}^\dag(\vec{k},\vec{q},0)e^{\imath\omega_{\vec{k}}t-i\vec{q}\cdot\vec{r}}]
\vec{s}_{\nu}(\vec{q}).\nonumber\\
\end{eqnarray}
Therefore the structural equations (\ref{d8}), (\ref{d12}) and
(\ref{d17}) together with the Maxwell equations, can be obtained
directly from the Heisenberg equations applied to the
electromagnetic field and the reservoirs.\\
For a homogeneous dielectric, the coupling functions $
f(\vec{r},\omega_{\vec{k}})$ and $ g(\vec{r},\omega_{\vec{k}}) $
are independent of $\vec{r}$, in this case from (\ref{d13}) and
(\ref{d19}), we deduce that $\chi_e$ and $\chi_m $ are
independent of position vector $\vec{r}$, hence the equation
(\ref{d9}) can be written as
\begin{eqnarray}\label{d23}
&&-\nabla^2\vec{A}+\frac{1}{c^2}\frac{\partial^2\vec{A}}{\partial
t^2}\pm \frac{1}{c^2}\frac{\partial}{\partial t}\int_0^{|t|}
dt'\chi_e(|t|-t')\frac{\partial \vec{A}}{\partial
t'}(\vec{r},\pm t')-\nonumber\\
&&\nabla\times\int_0^{|t|} dt'\chi_m(|t|-t')\nabla\times
\vec{A}(\vec{r},\pm t')
=\mu_0\frac{\partial\vec{P}^\bot_N}{\partial
t}(\vec{r},t)+\mu_0\nabla\times\vec{M}^\bot_N(\vec{r},t),\nonumber\\
&&
\end{eqnarray}
where the upper (lower) sign corresponds to  $t>0$ ($t<0$)
respectively.
\section{The solution of Heisenberg equations}
Let us denote the fourier transform of $\vec{A}(\vec{r},t) $ by $
\underline{\vec{A}}(\vec{q},t) $
\begin{equation}\label{d24}
\vec{A}(\vec{r},t)=\frac{1}{\sqrt{(2\pi)^3}}\int
d^3\vec{q}\underline{\vec{A}}(\vec{q},t)e^{i\vec{q}\cdot\vec{r}}
\end{equation}
from (\ref{d1}), it is clear that
\begin{equation}\label{d26}
\underline{\vec{A}}(\vec{q},t)=\sum_{\lambda=1}^2
\sqrt{\frac{\hbar} {2
\varepsilon_0\omega_{\vec{q}}}}[a_{\vec{q}\lambda}(t)\vec{e}_{\vec{q}\lambda}+a_{-\vec{q}
\lambda}^\dag(t)\vec{e}_{-\vec{q}\lambda}].
\end{equation}
The wave equation (\ref{d23}) can be written in terms of $
\underline{\vec{A}}(\vec{q},t)$
\begin{eqnarray}\label{d28}
&&\underline{\ddot{\vec{A}}}+\omega_{\vec{q}}^2\underline{\vec{A}}
\pm\frac{\partial}{\partial t}\int_0^{|t|}
dt'\chi_e(|t|-t')\dot{\underline{\vec{A}}}(\vec{q},\pm
t')-\omega_{\vec{q}}^2\int_0^{|t|}
dt'\chi_m(|t|-t')\underline{\vec{A}}(\vec{q},\pm t')\nonumber\\
&&=-\frac{\imath }{\varepsilon_0} \sum_{\lambda=1}^2 \int\frac{
d^3\vec{k}}{\sqrt{(2\pi)^3}}[\omega_{\vec{k}}f(\omega_{\vec{k}})d_{\lambda}(\vec{k},\vec{q},0)
e^{-\imath\omega_{\vec{k}}t}\vec{e}_{\vec{q}\lambda}-\omega_{\vec{k}}f^*(\omega_{\vec{k}})
d_{\lambda}^\dag(\vec{k},-\vec{q},0)e^{\imath\omega_{\vec{k}}t}\vec{e}_{-\vec{q}\lambda}]
\nonumber\\
&+&\omega_{\vec{q}}\sqrt{\mu_0}\sum_{\lambda=1}^2\int
\frac{d^3\vec{k}}{\sqrt{(2\pi)^3}}[g(\omega_{\vec{k}})b_{\lambda}(\vec{k},\vec{q},0)e^{-\imath\omega_{\vec{k}}t}\vec{e}_{\vec{q}\lambda}+
g^*(\omega_{\vec{k}})b_{\lambda}^\dag(\vec{k},-\vec{q},0)e^{\imath\omega_{\vec{k}}t}\vec{e}_{-\vec{q}\lambda}],\nonumber\\
&&
\end{eqnarray}
where $\omega_{\vec{q}}=c|\vec{q}|$. \\
This equation  can be solved using the Laplace transformation
method. For any time dependent operator $g(t)$, the forward and
backward Laplace transformations are by definition
\begin{equation}\label{d28.1}
g_f(s)=\int_0^\infty dt g(t)e^{-st},
\end{equation}
 and
\begin{equation}\label{d28.2}
g_b(s)=\int_0^\infty dt g(-t)e^{-st},
\end{equation}
respectively. One can solve  equation (\ref{d28}) using the
Laplace transformation. Let $ \tilde{\chi}_e(s)$ and
$\tilde{\chi}_m(s)$ be the Laplace transformations of $ \chi_e(t)
$ and $\chi_m(t)$ respectively, then $
\underline{\vec{A}}_f(\vec{q},s) $ and $
\underline{\vec{A}}_b(\vec{q},s) $, i.e., the forward and
backward Laplace transformation of $\vec{A}(\vec{q},t) $, can be
obtained in terms of $\tilde{\chi}_e(s)$ and $\tilde{\chi}_m(s)$
by taking the Laplace transformation of  equation  (\ref{d28})
\begin{eqnarray}\label{d29}
&&\underline{\vec{A}}_{f,b}(\vec{q},s)=\frac{s+s\tilde{\chi}_e(s)}{s^2+\omega_{\vec{q}}^2+s^2\tilde{\chi}_e(s)-\omega_{\vec{q}}^2\tilde{\chi}_m(s)}\underline{\vec{A}}(\vec{q},0)
\pm\frac{1}{s^2+\omega_{\vec{q}}^2+s^2\tilde{\chi}_e(s)-\omega_{\vec{q}}^2\tilde{\chi}_m(s)}\dot{\underline{\vec{A}}}(\vec{q},0)\nonumber\\
&&-\frac{\imath}{\varepsilon_0}\sum_{\lambda=1}^2\int
d^3\vec{k}\{\frac{\omega_{\vec{k}}f(\omega_{\vec{k}})d_\lambda(\vec{k},\vec{q},0)}{(s\pm\imath\omega_{\vec{k}})[s^2+\omega_{\vec{q}}^2+s^2\tilde{\chi}_e(s)-\omega_{\vec{q}}^2\tilde{\chi}_m(s)]}\vec{e}_{\vec{q}\lambda}\nonumber\\
&&-\frac{\omega_{\vec{k}}f^*(\omega_{\vec{k}})d_\lambda^\dag(\vec{k},-\vec{q},0)}{(s\mp\imath\omega_{\vec{k}})[s^2+\omega_{\vec{q}}^2+s^2\tilde{\chi}_e(s)-\omega_{\vec{q}}^2\tilde{\chi}_m(s)]}\vec{e}_{-\vec{q}\lambda}\}\nonumber\\
&&+\sqrt{\frac{\mu_0}{\varepsilon_0}}\omega_{\vec{q}}\sum_{\lambda=1}^2\int
d^3\vec{k}\{\frac{g(\omega_{\vec{k}})b_\lambda(\vec{k},\vec{q},0)}{(s\pm\imath\omega_{\vec{k}})[s^2+\omega_{\vec{q}}^2+s^2\tilde{\chi}_e(s)-\omega_{\vec{q}}^2\tilde{\chi}_m(s)]}\vec{e}_{\vec{q}\lambda}\nonumber\\
&&+\frac{g^*(\omega_{\vec{k}})b_\lambda^\dag(\vec{k},-\vec{q},0)}{(s\mp\imath\omega_{\vec{k}})[s^2+\omega_{\vec{q}}^2+s^2\tilde{\chi}_e(s)-\omega_{\vec{q}}^2\tilde{\chi}_m(s)]}\vec{e}_{-\vec{q}\lambda}\}
\end{eqnarray}
where the upper (lower) sign corresponds to $
\underline{\vec{A}}_f(\vec{q},s)$
($\underline{\vec{A}}_b(\vec{q},s)$) respectively. Now by taking
the inverse Laplace transformation of $
\underline{\vec{A}}_f(\vec{q},s) $ and $
\underline{\vec{A}}_b(\vec{q},s)$,
\begin{eqnarray}\label{d31}
&&\vec{A}(\vec{q},t)=r_{\pm}(\omega_{\vec{q}},t)\underline{\vec{A}}(\vec{q},0)\pm
h_{\pm}(\omega_{\vec{q}},t)\dot{\underline{\vec{A}}}(\vec{q},0)\nonumber\\
&&-\frac{\imath}{\varepsilon_0}\sum_{\lambda=1}^2\int
d^3\vec{k}\{\omega_{\vec{k}}f(\omega_{\vec{k}})\xi_{\pm}(\omega_{\vec{q}},\omega_{\vec{k}},t)d_\lambda(\vec{k},\vec{q},0)\vec{e}_{\vec{q}\lambda}-\omega_{\vec{k}}f^*(\omega_{\vec{k}})\xi_{\pm}^*(\omega_{\vec{q}},\omega_{\vec{k}},t)d_\lambda^\dag(\vec{k},-\vec{q},0)\vec{e}_{-\vec{q}\lambda}\}\nonumber\\
&&+\sqrt{\frac{\mu_0}{\varepsilon_0}}\omega_{\vec{q}}\sum_{\lambda=1}^2\int
d^3\vec{k}\{g(\omega_{\vec{k}})\xi_{\pm}(\omega_{\vec{q}},\omega_{\vec{k}},t)b_\lambda(\vec{k},\vec{q},0)\vec{e}_{\vec{q}\lambda}+g^*(\omega_{\vec{k}})\xi_{\pm}^*(\omega_{\vec{q}},\omega_{\vec{k}},t)b_\lambda^\dag(\vec{k},-\vec{q},0)\vec{e}_{-\vec{q}\lambda}\}\nonumber\\
&&
\end{eqnarray}
where
\begin{eqnarray}\label{d32}
&&r_{\pm}(\omega_{\vec{q}},t)=SOR\{\frac{(s+s\tilde{\chi}_e(s))e^{\pm
st}
}{s^2+\omega_{\vec{q}}^2+s^2\tilde{\chi}_e(s)-\omega_{\vec{q}}^2\tilde{\chi}_m(s)}\}\nonumber\\
&&h_{\pm}(\omega_{\vec{q}},t)=SOR\{\frac{e^{\pm st}
}{s^2+\omega_{\vec{q}}^2+s^2\tilde{\chi}_e(s)-\omega_{\vec{q}}^2\tilde{\chi}_m(s)}\}\nonumber\\
&&\xi_{\pm}(\omega_{\vec{q}},\omega_{\vec{k}},t)=SOR\{\frac{e^{\pm
st}
}{(s\pm \imath \omega_{\vec{k}})[s^2+\omega_{\vec{q}}^2+s^2\tilde{\chi}_e(s)
-\omega_{\vec{q}}^2\tilde{\chi}_m(s)]}\},\nonumber\\
&&
\end{eqnarray}
where the upper (lower) sign corresponds to $t>0$ $(t,0)$
respectively. $SOR(f(s))$ means the inverse Laplace
transformation of the complex $f(s)$ or simply the sum of residues
of $f(s)$.\\
The sum of the first and second term in (\ref{d31}) is the
solution of homogeneous  equation
\begin{equation}\label{d33}
\underline{\ddot{\vec{A}}}+\omega_{\vec{q}}^2\underline{\vec{A}}
\pm\frac{\partial}{\partial t}\int_0^{|t|}
dt'\chi_e(|t|-t')\dot{\underline{\vec{A}}}(\vec{q},\pm
t')-\omega_{\vec{q}}^2\int_0^{|t|}
dt'\chi_m(|t|-t')\underline{\vec{A}}(\vec{q},\pm t')=0
\end{equation}
and usually tends to zero in the limit $ t\rightarrow\pm\infty $
for the case of absorbing dielectrics. In this case if the
functions
\begin{eqnarray}\label{d34}
&&F(\omega_{\vec{q}},s)=\frac{s+s\tilde{\chi}_e(s)}{s^2+\omega_{\vec{q}}^2+s^2
\tilde{\chi}_e(s)-\omega_{\vec{q}}^2\tilde{\chi}_m(s)},\nonumber\\
&&G(\omega_{\vec{q}},s)=\frac{1}{s^2+\omega_{\vec{q}}^2+s^2\tilde{\chi}_e(s)-\omega_{\vec{q}}^2
\tilde{\chi}_m(s)},
\end{eqnarray}
have poles with negative real parts, then for enough large times,
\begin{eqnarray}\label{d35}
&&\vec{A}(\vec{r},t)=-\frac{\imath}{\varepsilon_0}\sum_{\lambda=1}^2\int
\frac{d^3\vec{q}}{\sqrt{(2\pi)^3}}\int
d^3\vec{k}\{\frac{\omega_{\vec{k}}
f(\omega_{\vec{k}})e^{-\imath\omega_{\vec{k}}t+\imath\vec{q}\cdot\vec{r}}}
{\omega_{\vec{q}}^2-\omega_{\vec{k}}^2-\omega_{\vec{k}}^2\tilde{\chi}_e(\mp\imath
\omega_{\vec{k}})-\omega_{\vec{q}}^2\tilde{\chi}_m(\mp\imath\omega_{\vec{k}})}
d_\lambda(\vec{k},\vec{q},0)\nonumber\\
&&-\frac{\omega_{\vec{k}}
f^*(\omega_{\vec{k}})e^{\imath\omega_{\vec{k}}t-\imath\vec{q}\cdot\vec{r}}}
{\omega_{\vec{q}}^2-\omega_{\vec{k}}^2-\omega_{\vec{k}}^2\tilde{\chi}_e(\pm\imath
\omega_{\vec{k}})-\omega_{\vec{q}}^2\tilde{\chi}_m(\pm\imath\omega_{\vec{k}})}
d_\lambda^\dag(\vec{k},\vec{q},0)
\}\vec{e}_{\vec{q}\lambda}+\nonumber\\
&&\sqrt{\frac{\mu_0}{\varepsilon_0}}\int\frac{d^3\vec{q}\omega_{\vec{q}}}{\sqrt{(2\pi)^3}}\int
d^3\vec{k}\{\frac{
g(\omega_{\vec{k}})e^{-\imath\omega_{\vec{k}}t+\imath\vec{q}\cdot\vec{r}}}
{\omega_{\vec{q}}^2-\omega_{\vec{k}}^2-\omega_{\vec{k}}^2\tilde{\chi}_e(\mp\imath
\omega_{\vec{k}})-\omega_{\vec{q}}^2\tilde{\chi}_m(\mp\imath\omega_{\vec{k}})}
b_\lambda(\vec{k},\vec{q},0)\nonumber+\\
&&\frac{g^*(\omega_{\vec{k}})e^{\imath\omega_{\vec{k}}t-\imath\vec{q}\cdot\vec{r}}}
{\omega_{\vec{q}}^2-\omega_{\vec{k}}^2-\omega_{\vec{k}}^2\tilde{\chi}_e(\pm\imath
\omega_{\vec{k}})-\omega_{\vec{q}}^2\tilde{\chi}_m(\pm\imath\omega_{\vec{k}})}
b_\lambda^\dag(\vec{k},\vec{q},0)\}\vec{e}_{\vec{q}\lambda},\nonumber\\
&&
\end{eqnarray}
However the complete solution for $ \vec{A}(\vec{r},t) $ can be
obtained by substitutiing $\underline{\vec{A}}(\vec{q},t)$ from
(\ref{d31}) in (\ref{d24}) and using (\ref{d6})
\begin{eqnarray}\label{d35.1}
&&\vec{A}(\vec{r},t)=\sum_{\lambda=1}^2\int
d^3\vec{q}\sqrt{\frac{\hbar}{2(2\pi)^3\varepsilon_0\omega_{\vec{q}}}}[Z_{\pm}
(\omega_{\vec{q}},t)e^{\imath\vec{q}\cdot\vec{r}}a_{\vec{q}\lambda}(0)+
Z^*_{\pm}(\omega_{\vec{q}},t)e^{-\imath\vec{q}\cdot\vec{r}}a_{\vec{q}\lambda}^\dag(0)]
\vec{e}_{\vec{q}\lambda}\nonumber\\
&&\pm\frac{1}{\varepsilon_0}\sum_{\lambda=1}^2\int
\frac{d^3\vec{q}}{\sqrt{(2\pi)^3}}\int
d^3\vec{k}\{\zeta_{\pm}(\omega_{\vec{k}},\omega_{\vec{q}},t)d_\lambda(\vec{k},\vec{q},0)
e^{\imath\vec{q}\cdot\vec{r}}+\zeta_{\pm}^*(\omega_{\vec{k}},\omega_{\vec{q}},t)
d_\lambda^\dag(\vec{k},\vec{q},0)e^{-\imath\vec{q}\cdot\vec{r}}\}\vec{e}_{\vec{q}\lambda}\nonumber\\
&&+\sqrt{\frac{\mu_0}{\varepsilon_0}}\sum_{\lambda=1}^2\int
\frac{d^3\vec{q}\omega_{\vec{q}}}{\sqrt{(2\pi)^3}}\int
d^3\vec{k}\{\eta_{\pm}(\omega_{\vec{k}},\omega_{\vec{q}},t)b_\lambda(\vec{k},\vec{q},0)
e^{\imath\vec{q}\cdot\vec{r}}+\eta_{\pm}^*(\omega_{\vec{k}},\omega_{\vec{q}},t)
b_\lambda^\dag(\vec{k},\vec{q},0)e^{-\imath\vec{q}\cdot\vec{r}}\}\vec{e}_{\vec{q}\lambda},\nonumber\\
&&
\end{eqnarray}
where
\begin{eqnarray}\label{d36}
&&Z_{\pm}(\omega_{\vec{q}},t)=SOR\{\frac{[s+s\tilde{\chi}_e(s)\mp\imath\omega_{\vec{q}}]
e^{\pm st}}{s^2+\omega_{\vec{q}}^2+s^2\tilde{\chi}_e(s)-\omega_{\vec{q}}^2\tilde{\chi}_m(s)}\}
\nonumber\\
&&\zeta_{\pm}(\omega_{\vec{k}},\omega_{\vec{q}},t)=f(\omega_{\vec{k}})SOR\{\frac{se^{\pm
st}}{(s\pm\imath\omega_{\vec{k}})[s^2+\omega_{\vec{q}}^2+s^2\tilde{\chi}_e(s)-
\omega_{\vec{q}}^2\tilde{\chi}_m(s)]}\}\nonumber\\
&&\eta_{\pm}(\omega_{\vec{k}},\omega_{\vec{q}},t)=g(\omega_{\vec{k}})SOR\{\frac{e^{\pm
st}}{(s\pm\imath\omega_{\vec{k}})[s^2+\omega_{\vec{q}}^2+s^2\tilde{\chi}_e(s)-
\omega_{\vec{q}}^2\tilde{\chi}_m(s)]}\}.\nonumber\\
&&
\end{eqnarray}
The transverse component of the electrical field is obtained from
$ \vec{E}^\bot =-\frac{\partial\vec{A}}{\partial t} $ .  Having
the vector potential $ \vec{A} $ we can easily obtain the
transverse component of electrical polarization density $
\vec{P}^\bot $ and magnetic polarization density $\vec{M}$ using
 (\ref{d12})
and (\ref{d17}).\\
 Taking the Laplace transformation of the
structural equation (\ref{d12}), the longitudinal component of
the electrical field can be obtained as
\begin{eqnarray}\label{d37}
\vec{E}^\|&=&-\frac{\vec{P}^\|}{\varepsilon_0}=-\frac{1}{\varepsilon_0}\int
\frac{d^3\vec{q}}{\sqrt{(2\pi)^3}}
\int
d^3\vec{k}[Q_\pm(\omega_{\vec{k}},t)f(\omega_{\vec{k}})d_3(\vec{k},\vec{q},0)
e^{i\vec{q}\cdot
\vec{r}}+\nonumber\\
&+&Q_\pm^*(\omega_{\vec{k}},t)f^*(\omega_{\vec{k}})d_3^\dag(\vec{k},\vec{q},0)
e^{-i\vec{q}\cdot
\vec{r}}]\hat{q}
\end{eqnarray}
where
\begin{equation}\label{d38}
Q_\pm(\omega_{\vec{k}},t)= SOR \{\frac{e^{\pm
st}}{(1+\tilde{\chi}_e(s))(s\pm i\omega_{\vec{k}})}\}.
\end{equation}
In the following we consider some important examples.\\
\textbf{Example 1:}\\
Let us assume $ f( \omega_{\vec{k}})=g(\omega_{\vec{k}})=0 $,
then from (\ref{d13}) and (\ref{d19}) we have $
\chi_e(t)=\chi_m(t)=0 $, and from (\ref{d36}) it is clear
\begin{equation}\label{d39}
Z_+(\omega_{\vec{q}},t)=Z_-(\omega_{\vec{q}},t)=e^{-i\omega_{\vec{q}}t},\hspace{1.50
cm } \eta_\pm=\zeta_\pm=0,
\end{equation}
therefore in this case, quantization of electromagnetic field is
reduced to the usual quantization
in the vacuum, a feature that can not be seen in some of previous models.\\
\textbf{Example 2 :}\\
Take $\chi_e(t)$ and $\chi_m(t)$ as follows
\begin{displaymath}\label{d40}
\chi_e(t)=\left\{\begin{array}{ll}
\frac{\chi_e^0}{\triangle} & 0<t<\triangle,\\
0 & \textrm{otherwise},
\end{array}\right.
\end{displaymath}
\begin{displaymath}\label{d41}
\chi_m(t)=\left\{\begin{array}{ll}
\frac{\chi_m^0}{(\chi_m^0+1)\triangle} & 0<t<\triangle\\
0 & \textrm{otherwise}
\end{array} \right.
\end{displaymath}
\begin{equation}\label{d42}
\end{equation}
where $\chi_e^0$, $\chi_m^0$ and $\triangle $ are some positive
constants, using (\ref{d14}) and (\ref{d20}) we find
\begin{eqnarray}\label{d42.1}
|f(\omega_{\vec{k}}|^2=\frac{\hbar
c^3\varepsilon_0\chi_e^0}{4\pi^2
\omega_{\vec{k}}^2}\frac{\sin^2\frac{\omega_{\vec{k}}\triangle}{2}}
{\frac{\omega_{\vec{k}}\triangle}{2}},\nonumber\\
|g(\omega_{\vec{k}}|^2=\frac{\hbar
c^3\chi_m^0}{4\pi^2\mu_0(\chi_m^0+1)
\omega_{\vec{k}}^2}\frac{\sin^2\frac{\omega_{\vec{k}}\triangle}{2}}
{\frac{\omega_{\vec{k}}\triangle}{2}},\nonumber\\
\end{eqnarray}
and from (\ref{d12}) and (\ref{d17}) we have
\begin{eqnarray}\label{d42.2}
&&\vec{P}(\vec{r},t)=\vec{P}_N(\vec{r},t)+\frac{\varepsilon_0\chi_e^0}{\triangle}
\int_{|t|-\triangle}^|t| d t'\vec{E}(\vec{r},\pm t'),\nonumber\\
&&\vec{M}(\vec{r},t)=\vec{M}_N(\vec{r},t)+\frac{\chi_m^0}{\mu_0(\chi_m^0+1)
\triangle}\int_{|t|-\triangle}^|t| d t'\vec{B}(\vec{r},\pm t'),
\end{eqnarray}
where $\vec{P}_N(\vec{r},t),\vec{M}_N(\vec{r},t)$ are the noise
polarization densities corresponding to the coupling
functions obtained in (\ref{d42.1}).\\
In the limit $ \triangle \rightarrow0 $, the coupling functions
(\ref{d42.1}) tend to zero, and the relations (\ref{d42.2}) are
reduced to
\begin{eqnarray}\label{d42.3}
\vec{P}(\vec{r},t)&=&\varepsilon_0\chi_e^0\vec{E}(\vec{r},t),\nonumber\\
\vec{M}(\vec{r},t)&=&\frac{\chi_m^0}{\mu_0(\chi_m^0+1)}\vec{B}(\vec{r},t).\nonumber\\
\end{eqnarray}
In this limit the electrical field and the polarization densities
are purely transverse and
\begin{eqnarray}\label{d42.3}
&&
Z_\pm(\omega_{\vec{q}},t)=\cos\tilde{\omega}_{\vec{q}}t-i\sqrt{\frac{1+\chi_m^0}{1+\chi_e^0}}
\sin\tilde{\omega}_{\vec{q}}t,\hspace{1.00 cm}
\tilde{\omega}_{\vec{q}}=\frac{\omega_{\vec{q}}}{\sqrt{(1+\chi_e^0)(1+\chi_m^0)}},\nonumber\\
&&\eta_\pm(\omega_{\vec{q}},t)=\zeta_\pm(\omega_{\vec{q}},t)=0.
\end{eqnarray}
  The electromagnetic energy inside the
dielectric is
\begin{eqnarray}\label{d45}
\int[\frac{1}{2}\vec{E}.\vec{D}+\frac{1}{2}\vec{H}.\vec{B}]d^3r&=&
\int d^3\vec{q}[\frac{1}{2\varepsilon_0(1+\chi_e^0)}
\underline{\vec{D}}(\vec{q}.0)\cdot
\underline{\vec{D}}^\dag(\vec{q}.0)+\nonumber\\
&+&\frac{\varepsilon_0\omega_{\vec{q}}^2}{2(1+\chi_m^0)}\underline{\vec{A}}(\vec{q},0)\cdot
\underline{\vec{A}}^\dag(\vec{q},0)]
\end{eqnarray}
where $\underline{\vec{D}} $ is the Fourier transform of the
displacement field, this energy ia a constant of motion contrary
to the vacuum expression
$\int[\frac{1}{2}\varepsilon_0\vec{E}^2+\frac{\vec{B}^2}{2\mu_0}]d^3r$,
which is not a constant of motion.\\
\textbf{Example 3:}
 Let $\chi_e(t)=\beta u(t)$ and $\chi_m(t)=0$ where $ u(t) $ is
 the step function
 \begin{displaymath}
u(t)=\left\{\begin{array}{ll}
1 & t\leq 0\\
0 & t>0
\end{array}\right.
\end{displaymath}
\begin{equation}\label{d46}
\end{equation}
and $ \beta $ is a positive constant, then using (\ref{d14}) and
(\ref{d20}) we find
\begin{equation}\label{d47}
|f(\omega_{\vec{k}}|^2=\frac{\hbar
c^3\varepsilon_0\beta}{4\pi^2\omega_{\vec{k}}^3},\hspace{2.00
cm}g(\omega_{\vec{k}})=0,
\end{equation}
and accordingly we can rewrite (\ref{d28}) as
\begin{eqnarray}\label{d48}
\ddot{\underline{\vec{A}}}+\omega_{\vec{q}}^2 \underline{\vec{A}}
+\beta\dot{\underline{\vec{A}}}&=& -\imath\sqrt{\frac{\hbar
c^3\beta}{4\pi^2 \varepsilon_0}}
\int\frac{d^3\vec{k}}{\sqrt{(2\pi)^3\omega_{\vec{k}}}}\sum_{\lambda=1}^2[d_{\lambda}
(\vec{k},\vec{q},0)e^{-\imath\omega_{\vec{k}}t}\vec{e}_{\vec{q}\lambda}\nonumber\\
&-&d_{\lambda}^\dag(\vec{k},-\vec{q},0)e^{\imath
\omega_{\vec{k}}t}\vec{e}_{-\vec{q}\lambda}],
\end{eqnarray}
which has a dissipative term proportional to velocity. From
(\ref{d36}) one can obtain
\begin{eqnarray}\label{d48.1}
&&Z_\pm(\omega_{\vec{q}},t)=e^{\mp\frac{\beta}{2}t}[\pm\frac{\beta}{2\Omega_{\vec{q}}}\sin\Omega_{\vec{q}}t+
\cos\Omega_{\vec{q}}t-\frac{i\omega_{\vec{q}}}{\Omega_{\vec{q}}}\sin\Omega_{\vec{q}}t],\nonumber\\
&&\zeta_\pm(\omega_{\vec{k}},\omega_{\vec{q}},t)=\sqrt{\frac{\hbar
c^3\beta\varepsilon_0}{4\pi^2\omega_{\vec{k}}^3}}\{\mp\frac{i\omega_{\vec{k}}e^{-i\omega_{\vec{k}}t}}{\omega_{\vec{q}}^2-\omega_{\vec{k}}^2\mp
i\beta\omega_{\vec{k}}}+\nonumber\\
&&e^{\mp\frac{\beta}{2}t}[\frac{(-\frac{\beta}{2}+i\Omega_{\vec{q}})e^{\pm
i\Omega_{\vec{q}}t}}{2i\Omega_{\vec{q}}(-\frac{\beta}{2}+i\Omega_{\vec{q}}\pm
i\omega_{\vec{k}}
)}+\frac{(\frac{\beta}{2}+i\Omega_{\vec{q}})e^{\mp
i\Omega_{\vec{q}}t}}{2i\Omega_{\vec{q}}(-\frac{\beta}{2}-i\Omega_{\vec{q}}\pm
i\omega_{\vec{k}} )}\},
\end{eqnarray}
where $
\Omega_{\vec{q}}=\sqrt{\omega_{\vec{q}}^2-\frac{\beta^2}{4}} $.
Therefore
 the stable solution of $\vec{A}(\vec{r},t) $ in the limit $t\rightarrow
\pm\infty$, is
\begin{eqnarray}\label{d49}
&&\vec{A}(\vec{r},t)=\mp\imath\sqrt{\frac{\hbar c^3\beta}{4\pi^2
\varepsilon_0 }}\sum_{\lambda=1}^2
\int\frac{d^3\vec{q}}{\sqrt{(2\pi)^3}}
\int\frac{d^3\vec{k}}{\sqrt{\omega_{\vec{k}}}}[\frac{d_{\lambda}
(\vec{k},\vec{q},0)e^{-\imath\omega_{\vec{k}}t+i\vec{q}\cdot\vec{r}}}{\omega_{\vec{q}}^2-\omega_{\vec{k}}^2\mp\imath\beta
\omega_{\vec{k}}}-\nonumber\\
&&\frac{d_{\lambda}^\dag(\vec{k},\vec{q},0)e^{\imath\omega_{\vec{k}}t-i\vec{q}\cdot\vec{r}}}
{\omega_{\vec{q}}^2-\omega_{\vec{k}}^2\pm\imath\beta\omega_{\vec{k}}}]\vec{e}_{\vec{q}\lambda}.
\end{eqnarray}
from (\ref{d37}) it is easy to show that the longitudinal
component of the electrical field in the limit $t\rightarrow
\pm\infty$ is
\begin{eqnarray}\label{d49.1}
\vec{E}^\|=-\frac{\vec{P}^\|}{\varepsilon_0}&=&\pm i\sqrt{
\frac{\hbar
c^3\beta}{4\pi^2\varepsilon_0}}\int\frac{d^3\vec{q}}{\sqrt{(2\pi)^3}}
\int
\frac{d^3\vec{k}}{\sqrt{\omega_{\vec{k}}}}[\frac{d_3(\vec{k},\vec{q},0)}{\beta\mp
i\omega_{\vec{k}} }e^{-i\omega_{\vec{k}}t+i\vec{q}\cdot
\vec{r}}\nonumber\\
&-&\frac{d_3^\dag(\vec{k},\vec{q},0)}{\beta\pm i\omega_{\vec{k}}
}e^{i\omega_{\vec{k}}t-i\vec{q}\cdot \vec{r}}]\hat{q}.
\end{eqnarray}
\textbf{Example 4: A simple model for $ \tilde{\chi}_e(s) $ }\\
If we neglect the difference between local and macroscopic
electric field for substances with a low density, then the
classical equation of a bound atomic electron in an external
electrical field is
\begin{equation}\label{d51}
\ddot{\vec{r}}+\gamma\dot{\vec{r}}+\omega_0^2\vec{r}=-\frac{e}{m}\vec{E}(t),
\end{equation}
where $\gamma$ is a damping coefficient and the force exerted on
the electron due to atom is taken to be simply a spring force
with frequency $ \omega_0 $ and the magnetic force has been
neglected in comparison with the electric one. If $
\tilde{\vec{E}}(s) $ and $ \tilde{\vec{r}}(s) $ are the Laplace
transformations of $ \vec{E}(t) $ and $ \vec{r}(t) $
respectively, then from (\ref{d51}) we can find
\begin{equation}\label{d52}
\tilde{\vec{r}}(s)
=\frac{-\frac{e}{m}\tilde{\vec{E}}(s)}{s^2+\gamma s+\omega_0^2},
\end{equation}
now let there be $ N $ molecules per unit volume with $ z $
electrons per molecule such that $ f_j $ electrons of any
molecule have a bound frequency $ \omega_j $ and a damping
coefficient $ \gamma_j $, then the Laplace transformation of the
polarization density is
\begin{equation}\label{d53}
\tilde{\vec{P}}(s)=\frac{Ne^2}{m}\sum_j\frac{f_j}{s^2+\gamma_js+\omega_j^2}
\tilde{\vec{E}}(s).
\end{equation}
If $ \omega_j $ and $ \gamma_j $ are identical for all of
electrons, then from (\ref{d53})
  \begin{eqnarray}\label{d54}
&&\tilde{\chi}_e(s)=\frac{\omega_p^2}{s^2+\gamma
s+\omega_0^2},\hspace{1.5
cm}\omega_p^2=\frac{Ne^2z}{m\varepsilon_0},\nonumber\\
&&\chi_e(t)=\omega_p^2e^{-\frac{\gamma
t}{2}}\frac{\sin\nu_0t}{\nu_0}u(t),\hspace{0.7
cm}\nu_0^2=\omega_0^2-\frac{\gamma^2}{4},
\end{eqnarray}
where $ u(t) $ is the step function defined in (\ref{d46}). We can
obtain the coupling function $ f(\omega_{\vec{k}}) $ from
(\ref{d14}) as
 \begin{eqnarray}\label{d55}
 |f(\omega_{\vec{k}})|^2=\frac{\hbar c^3\varepsilon_0\omega_p^2}{16\pi^2\nu_0
 \omega_{\vec{k}}^2}\{\frac{\gamma}{\frac{\gamma^2}{4}+(\nu_0-\omega_{\vec{k}})^2}-
 \frac{\gamma}{\frac{\gamma^2}{4}+(\nu_0+\omega_{\vec{k}})^2}\},
\end{eqnarray}
If $ \gamma=0 $, then the dielectric substance is a
nondissipative one and
\begin{equation}\label{d58}
|f(\omega_{\vec{k}})|^2=\frac{\hbar
c^3\varepsilon_0\omega_p^2}{8\pi\nu_0^3}\delta(\nu_0-\omega_{\vec{k}}).
\end{equation}
In this case the noise electrical polarization density is nonzero
only for frequency $ \omega_{\vec{k}}=\omega_0 $, because $
\omega_0 $ is the resonance frequency of equation $
\ddot{\vec{r}}+\omega_0^2 \vec{r}=-\frac{e}{m}\vec{E}_0
e^{-i\omega_0 t} $ and in this frequency the energy of
electromagnetic field is absorbed by the dielectric. From
(\ref{d36}) we have
\begin{equation}\label{d59}
Z_\pm(\omega_{\vec{q}},t) = SOR[\frac{(s\mp
i\omega_{\vec{q}})(s^2+\omega_0^2)+s\omega_p^2}{s^4+s^2(\omega_0^2+\omega_{\vec{q}}^2+\omega_p^2)
+\omega_{\vec{q}}^2\omega_0^2}e^{\pm st}]
\end{equation}
 which is the sum of residues of the function inside the bracket with poles
 at $\pm i\Omega_\pm$, where
 \begin{equation}\label{d60}
\Omega_\pm
=\frac{\omega_0^2+\omega_{\vec{q}}^2+\omega_p^2}{2}\mp\sqrt{\frac{\omega_0^2+
\omega_{\vec{q}}^2+\omega_p^2}{4}-\omega_0^2\omega_{\vec{q}}^2}.
\end{equation}
Similarly $\zeta_\pm$ can be obtained from (\ref{d36}) with $
f(\omega_{\vec{k}}) $ given in (\ref{d58}). In this case
\begin{equation}\label{d61}
Q_\pm(\omega_{\vec{k}},t)=\frac{\omega_0^2-\omega_{\vec{k}}^2}{\omega_0^2+
\omega_p^2-\omega_{\vec{k}}^2}e^{-i\omega_{\vec{k}}t}
+\frac{\omega_p^2}{2\sqrt{\omega_0^2+\omega_p^2}}\{\frac{e^{\pm
i\sqrt{\omega_0^2+\omega_p^2}t}}{\sqrt{\omega_0^2+\omega_p^2}\pm\omega_{\vec{k}}}+
\frac{e^{\mp
i\sqrt{\omega_0^2+\omega_p^2}t}}{\sqrt{\omega_0^2+\omega_p^2}
\mp\omega_{\vec{k}}}\},
\end{equation}
and the longitudinal component of the electrical field can be
obtained from (\ref{d37}) using $f(\omega_{\vec{k}})$ and $
Q_\pm(\omega_{\vec{k}},t)$ given by (\ref{d58}) and (\ref{d61})
respectively.\\
 If $\gamma\neq 0$, the substance is of
dissipative kind and $\vec{A}(\vec{r},t)$, in the limit $
t\rightarrow \pm\infty $ can be obtained from (\ref{d35}) with $
f(\omega_{\vec{k}}) $ given by (\ref{d55}) and $
\tilde{\chi}_e(\mp
i\omega_{\vec{k}})=\frac{\omega_p^2}{\omega_0^2-\omega_{\vec{k}}^2\pm
i\gamma \omega_{\vec{k}}}$. In this case in the limit $
t\rightarrow\pm\infty $ we have
\begin{equation}\label{d62}
Q_\pm(\omega_{\vec{k}},t) = \frac{\omega_0^2-\omega_{\vec{k}}^2\mp
i\gamma\omega_{\vec{k}}}{\omega_0^2+\omega_p^2-\omega_{\vec{k}}^2\mp
i\gamma \omega_{\vec{k}}}e^{-i\omega_{\vec{k}}t},
\end{equation}
and the longitudinal component of the electrical field can be
obtained from (\ref{d37}). This example shows that this model of
quantization of the electromagnetic field is applicable to both
dissipative and nondissipative dielectrics.
\section{Concluding remarks}
By modeling the dieletric medium with two independent reservoirs,
electric and magnetic reservoirs, we could investigate
consistently the electromagnetic field quantization inside a
linear dielectric medium. If a definite dielectric is given,
i.e., $\chi_e(t)$ and $\chi_m(t)$ are definite functions, then we
could find the corresponding coupling functions
$f(\omega_{\vec{k}})$ and $g(\omega_{\vec{k}})$, respectively for
the reservoirs. In this approach, both the Maxwell and structural
equations obtained together. In the limiting case, i.e., when
there is no dielectric, the approach tends to the usual
quantization of the electromagnetic field  in vacuum, a feature
that can not be seen in some of the previous models. Also this
model of quantization is applicable to both dissipative and
nondissipative dielectrics.

\end{document}